\documentclass[%
 reprint,
 amsmath,amssymb,
 aps,
floatfix,
]{revtex4-2}
\usepackage{multirow}
\usepackage{graphicx}
\usepackage{dcolumn}
\usepackage{bm}
\usepackage{hyperref}
\usepackage[mathlines]{lineno}

\usepackage[showframe],

\usepackage{siunitx}
\usepackage{xcolor}

\begin{document}

\preprint{APS/123-QED}

\title{Complex Antiferromagnetic Order in the \\Metallic Triangular Lattice Compound SmAuAl$_4$Ge$_2$} 

\author{Keke Feng}
\altaffiliation {Department of Physics, Florida State University.}
\altaffiliation {National High Magnetic Field Laboratory}

\author{Caleb Bush}
\altaffiliation {Department of Physics, Rochester Institute of Technology.}

\author{Olatunde Oladehin}
\altaffiliation {Department of Physics, Florida State University.}
\altaffiliation {National High Magnetic Field Laboratory}

\author{Minhyea Lee}
\altaffiliation {Department of Physics, University of Colorado, Boulder.}
\altaffiliation {National High Magnetic Field Laboratory}

\author{Ryan Baumbach}
\altaffiliation {Department of Physics, Florida State University.}
\altaffiliation {National High Magnetic Field Laboratory}

\date{\today}

\begin{abstract}
The compounds $Ln$AuAl$_4$Ge$_2$ ($Ln$ $=$ lanthanide) form in a structure that features two-dimensional triangular lattices of $Ln$ ions that are stacked along the crystalline $c$ axis. Together with crystal electric field effects, magnetic anisotropy, and electron-mediated spin exchange interactions, this sets the stage for the emergence of strongly correlated spin and electron phenomena. Here we investigate SmAuAl$_4$Ge$_2$, which exhibits weak paramagnetism that strongly deviates from conventional Curie-Weiss behavior. Complex antiferromagnetic ordering emerges at $T_{\rm{N1}}$ $=$ 13.2 K and $T_{\rm{N2}}$ $=$ 7.4 K, where heat capacity measurements show that these transitions are first and second order, respectively. These measurements also reveal that the Sommerfeld coefficient is not enhanced compared to the nonmagnetic analog YAuAl$_4$Ge$_2$, consistent with the charge carrier quasiparticles exhibiting typical Fermi liquid behavior. The temperature-dependent electrical resistivity follows standard metallic behavior, but linear magnetoresistance unexpectedly appears within the ordered state. We compare these results to other $Ln$AuAl$_4$Ge$_2$ materials, which have already been established as localized $f$-electron magnets that are hosts for interesting magnetic and electronic phases. From this, SmAuAl$_4$Ge$_2$ emerges as a complex quantum spin metal, inviting further investigations into its properties and the broader family of related materials.
\begin{description}
\item[PACS numbers]
PACS
\end{description}
\end{abstract}
\pacs{Valid PACS appear here}
\maketitle
\section{\label{sec:intro}Introduction}
Geometrically frustrated magnetic materials historically have attracted substantial interest as reservoirs for novel quantum phases~\cite{ramirez1,ramirez2,balents,zhou}. A prototypical example is the insulating two-dimensional triangular antiferromagnetic spin lattice, where early theoretical efforts indicated that the ground state for spin $S$ = 1/2 does not exhibit long-range order~\cite{anderson}. Later numerical studies modified this model by showing that 120$^o$ order can emerge when antiferromagnetic nearest neighbor spin interactions are included~\cite{AFM-120-1,AFM-120-2,AFM-120-3,AFM-120-4}. Since then, substantial experimental studies have been carried out for insulators with triangular nets of transition metal elements, many of which exhibit complex magnetic phenomena~\cite{Insulator-RuCl3-1,Insulator-RuCl3-2,Insulator-RuCl3-3,Insulator-Bi2Te3}. Efforts have also been extended to lanthanide-containing systems, where the $f$-electron wave function is well localized by comparison to $d$-electron analogs. This results in weakened exchange couplings and large $g$-factors, producing rich phase diagrams with easily tuned ground states: e.g., using magnetic fields~\cite{Ln-frustrated-SL1, CsYbSe2}. More recently, related semimetal and metal systems have also been studied, where the presence of conduction electrons leads to long-range magnetic exchange interactions that are mediated by the Ruderman-Kittel-Kasuya-Yosida (RKKY) mechanism\cite{RKKY1,RKKY2,RKKY3}. In addition to enhancing the degrees of freedom, this enables opportunities to explore the emergence of flat bands~\cite{kang20,sales21}, exotic superconductivity~\cite{ortiz1,ortiz2}, skyrmion phases~\cite{Gd2PdSi3}, unconventional Hall effects~\cite{Kagome,166}, and potentially even metallic quantum spin liquids~\cite{gegenwart}.

This motivated us to examine the compounds $Ln$$T$Al$_4$Ge$_2$ ($Ln$ = lanthanide and $T$ $=$ transition metal), which feature triangular nets of $Ln$ ions~\cite{1142} (Fig.~\ref{fig1}). Studies of the $Ln$ = Ce, Nd, Gd, and Tb variants have recently revealed complex ordering, indications of magnetic frustration, and complex temperature-magnetic field ($T-H$) phase diagrams. For example, CeAuAl$_4$Ge$_2$ hosts trivalent cerium moments with evidence for a ferromagnetic interaction within the triangular $ab$-plane, which might relieve spin frustration~\cite{CeAuAl4Ge2}. In the case of NdAuAl$_4$Ge$_2$, the spins prefer to orient along the $c$-axis, two magnetic phase transitions are observed ($T_{\rm{N1}}$ $=$ 1.75 K and $T_{\rm{N2}}$ $=$ 0.49 K), and several metamagnetic phase transitions are seen for magnetic fields applied along the $c$-axis~\cite{3NdAuAl4Ge2.PRM.2023}. Even more complex behavior is seen for GdAuAl$_4$Ge$_2$ and TbAuAl$_4$Ge$_2$, which exhibit multiple transitions at substantially higher ordering temperatures, easy-$ab$-plane anisotropy, and multiple anisotropic metamagnetic phase transitions for magnetic fields applied in the triangular net plane\cite{1LnAuAl4Ge2,2LnAuAl4Ge2}. Finally, it is noteworthy that CePtAl$_4$Ge$_2$ exhibits heavy fermion antiferromagnetism, showing that Kondo lattice physics impacts behavior in some cases~\cite{CePtAl4Ge2}. Thus, it is natural to anticipate that further variation of the $Ln$ ion could produce other novel phenomena resulting from the combination of crystal electric field effects, complex RKKY interactions, geometric frustration, and strong electronic correlations.

Here we focus on SmAuAl$_4$Ge$_2$, where the $f$-electron state is likely to be distinct from that of its lanthanide neighbors. In particular, samarium $f$-electrons can (i) adopt either a divalent (4$f^6$; $J$ $=$ 0) or trivalent (4$f^5$; $J$ $=$ 5/2) configuration, with crystal electric field splitting, (ii) hybridize with conduction electron states, and (iii) exhibit van Vleck paramagnetism due to the ground state ($J$ $=$ 5/2) and first excited state ($J$ $=$ 7/2) being separated by a relatively small energy~\cite{vleck,vanVlec2}.  This generates interesting behavior in other model systems, including (i) heavy fermion ground states in SmOs$_4$Sb$_{12}$ \cite{SmOs4Sb121,SmOs4Sb122}, SmPt$_4$Ge$_{12}$~\cite{SmPt4Ge12}, and SmTi$_2$Al$_{20}$ \cite{SmTi2Al20}; (ii) topological Kondo insulating behavior in SmB$_6$ \cite{menthe,SmB6}; (iii) coexistence of superconductivity and magnetism in SmRh$_4$B$_4$~\cite{SmRh4B4}; and (iv) valence instabilities in samarium monochalcogenides~\cite{jayaraman}. 

From magnetization, heat capacity, and electrical transport measurements, we show that SmAuAl$_4$Ge$_2$ exhibits strong deviations from conventional Curie-Weiss paramagnetism at elevated temperatures due to crystal electric field splitting of the $J$ $=$ 5/2 multiplet, and possibly other effects. Complex antiferromagnetic ordering appears at $T_{\rm{N1}}$ $=$ 13.2 K and $T_{\rm{N2}}$ $=$ 7.4 K, which are first and second order transitions, respectively. Metallic behavior without evidence for enhanced mass charge carrier quasiparticles is seen in the electronic coefficient of the heat capacity and electrical transport measurements, resembling what is seen for the nonmagnetic analog YAuAl$_4$Ge$_2$. Interestingly, although applied magnetic fields up to 9 T have little effect on the ordering temperatures, linear magnetoresistance resembling what is seen for GdAuAl$_4$Ge$_2$ and TbAuAl$_4$Ge$_2$~\cite{1LnAuAl4Ge2,2LnAuAl4Ge2} is observed within the ordered state. Thus, SmAuAl$_4$Ge$_2$ emerges as an intriguing addition to the $Ln$AuAl$_4$Ge$_2$ family, where complex magnetic ordering and unusual magnetotransport behavior are observed within an ensemble of $f$-electron spins whose high-temperature paramagnetism differs significantly from other $Ln$ analogs.
\section{\label{sec:methods}Experimental Methods}
SmAuAl$_4$Ge$_2$ single crystals were grown using an aluminum molten metal flux following the procedure detailed in Refs.~\cite{1142,CeAuAl4Ge2}. In order to allow comparison to a non-$f$-electron containing analogue, single crystal specimens of YAuAl$_4$Ge$_2$ were produced using the same method. Crystals typically form as three-dimensional clusters, where individual crystals with dimensions on the order of 2 mm and hexagonal or triangular facets associated with the $ab$ plane can be isolated (Fig.~\ref{fig1}). Room temperature powder x-ray diffraction (PXRD) measurements were performed using a Rigaku SmartLab SE X-ray diffractometer with a Cu K$\alpha$ source. The Rietveld refinement analysis was done using GSAS-II to assess the purity and determine the structure parameters. The principal c-axis was identified by measuring the diffraction pattern on polished flat crystals using the same system and was also apparent in the crystal shape. 

Temperature ($T$) dependent magnetization $M$ measurements were carried out for $T$ = 1.8 - 300 K under magnetic fields of $\mu_0 H$ = 0.5T applied parallel ($\|$) and perpendicular ($\perp$) to the crystallographic $c$-axis using a Quantum Design VSM Magnetic Property Measurement System. Data were collected (i) under zero field cooled (ZFC) conditions, where the sample was cooled to $T$ $=$ 1.8 K, the magnetic field was applied, and $M$ was measured as $T$ increased to 300 K and (ii) field cooled (FC) conditions, where the magnetic field was applied at 300 K and $M$ was measured as $T$ decreased to 1.8 K. Isothermal magnetization measurements were also performed for $\mu_0 H$ $\leq$ 7 T, where the sample was zero field cooled prior to the measurement at each temperature. Heat capacity $C$ measurements were performed for $T$ = 1.8 - 40 K in a Quantum Design Physical Properties Measurement System using a conventional thermal relaxation technique. Electrical resistivity $\rho$ measurements for $T$ = 1.8 - 300 K and magnetic fields $\mu_0 H$ $\leq$ 9 T were performed in a four-wire configuration for polished single crystal using the same system. For $\rho(T)$, both ZFC and FC measurements were performed. For $\rho(H)$, samples were zero field cooled prior to measurements.


\section{\label{sec:results}Results}

\begin{figure}
  \includegraphics[width=\columnwidth]{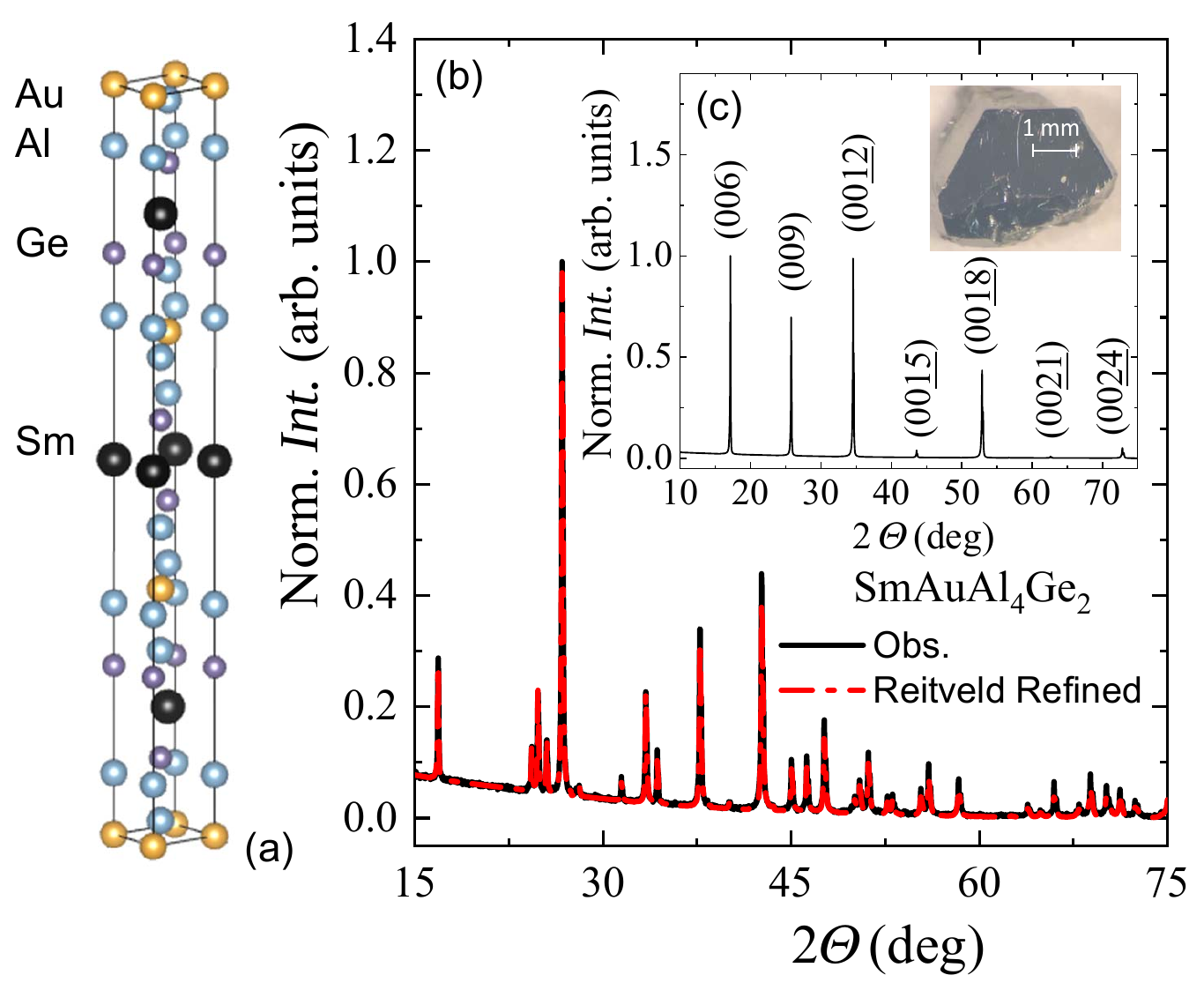}\caption{\label{fig1} (a) Crystal structure of SmAuAl$_4$Ge$_2$~\cite{vesta}. (b) Rietveld refinement of the powder X-ray diffraction pattern for SmAuAl$_4$Ge$_2$. The black line is the observed experimental pattern, and the red dashed line is the calculated pattern. Inset: A single crystal specimen obtained from the aluminum flux growth method described in the text. (c) XRD patterns of a $c$-axis aligned SmAuAl$_4$Ge$_2$ single crystal similar to that shown in the inset.}
\end{figure}

The trigonal SmAuAl$_4$Ge$_2$ unit cell is shown in Fig.~\ref{fig1}(a). The powder X-ray diffraction pattern for SmAuAl$_4$Ge$_2$ is shown in Fig.~\ref{fig1}(b), where the data are described by the trigonal $Ln$AuAl$_4$Ge$_2$ structure (space group $R\overline{3}m$ (No.~166)~\cite{1142}). A Rietveld refinement yields lattice parameters and a unit cell volume $a$ $=$ 4.21711(10) \AA, 31.1452(6) \AA,~ and $V$ $=$ 479.680(13), consistent with expectations for the trivalent lanthanide contraction for $Ln$AuAl$_4$Ge$_2$ discussed in Ref.~\cite{1LnAuAl4Ge2}. Other fit parameters are summarized in Table I. Fig.~\ref{fig1}(c) shows the XRD pattern for a $c$-axis aligned crystal of SmAuAl$_4$Ge$_2$, exhibiting only the (00$l$) diffraction peaks. This reveals that the naturally occurring hexagonal facets are aligned in the $ab$ plane.

\begin{table}
    \centering
    \begin{tabular}{|c|c|c|c|}\hline
        Site Label & $x, y, z$ & occupancy & U$_{\rm{iso}}$ \\\hline
        Al2  & 0, 0, 0.08089 & 1 & 0.0199 \\\hline
        Ge   & 0, 0, 0.22380 & 1 & 0.0136 \\\hline
        Al1  & 0, 0, 0.31009 & 1 & 0.0538 \\\hline
        Sm   & 0, 0, 0.50000 & 1 & 0.0027 \\\hline
        Au   & 0, 0, 0       & 1 & 0.0018 \\\hline
    \end{tabular}
    \caption{Summary of crystallographic parameters resulting from Rietveld refinement of the data using GSAS-II. Fits yielded the lattice constants $a$ $=$ 4.21711(10) \AA, 31.1452(6) \AA,~ and $V$ $=$ 479.680(13). The quality of the fit is characterized by $R_{\rm{W}}$ = 7.76.}
    \label{tab:my_label}
\end{table}



\begin{figure}
   \includegraphics[width=\columnwidth]{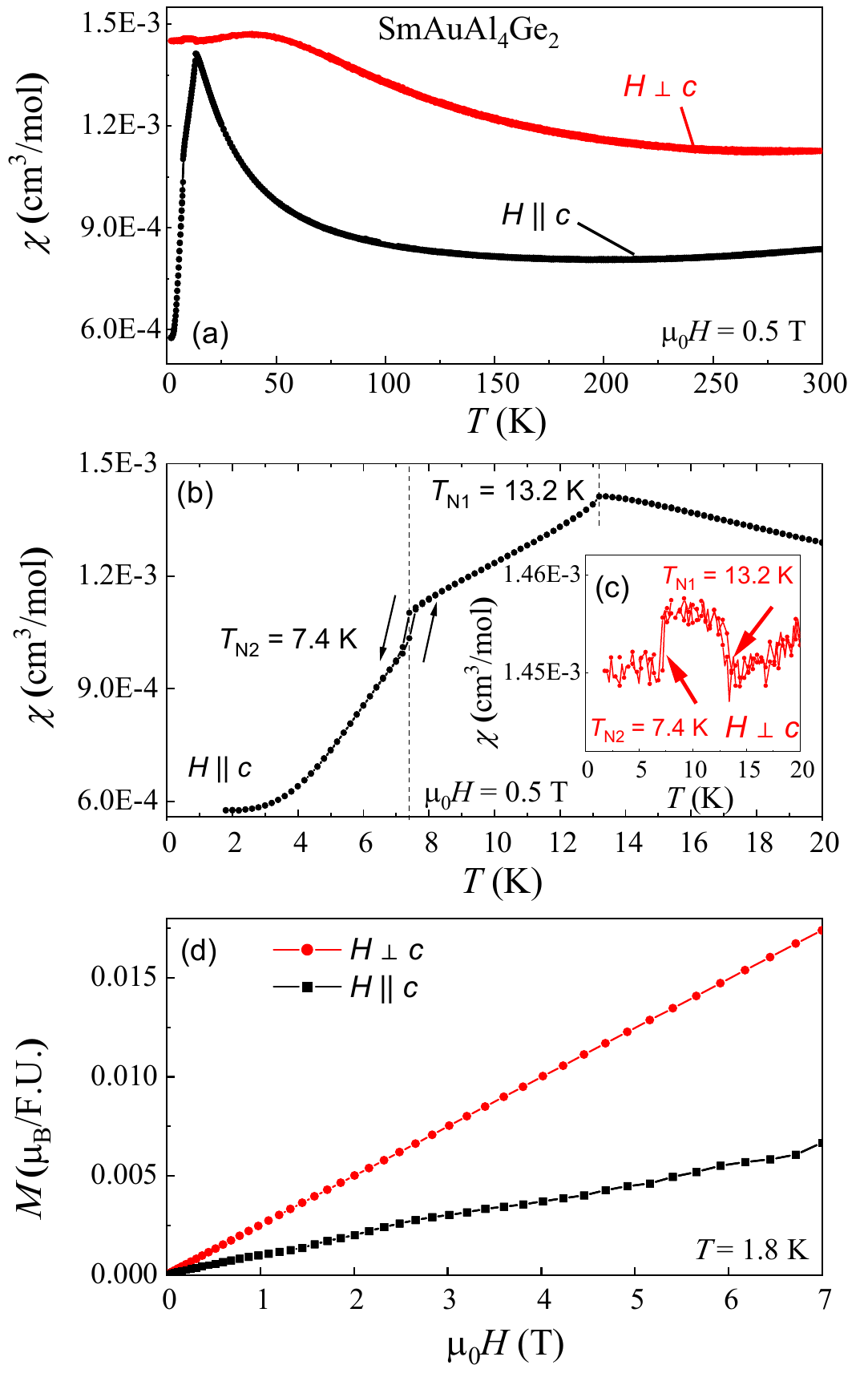}\caption{\label{fig2} (a) Temperature dependent magnetic susceptibility $\chi(T)$ for SmAuAl$_4$Ge$_2$ for magnetic fields $\mu_0H$ $=$ 0.5 T applied parallel ($\parallel$) and perpendicular ($\perp$) to the $c$ axis. Data were collected both for zero field cooling and field cooling, as described in the experimental methods. (b,c) Zoom of $\chi(T)$ at low temperatures for magnetic fields $\mu_0H$ $=$ 0.5 T applied (b) parallel ($\parallel$) and (c) perpendicular ($\perp$) to the $c$ axis, emphasizing the magnetic phase transitions. The dotted vertical lines in (b) represent the locations of the ordering temperatures $T_{\rm{N1}}$ and $T_{\rm{N2}}$. (d) Isothermal magnetization $M(H)$ for both field directions at $T$ $=$ 1.8 K.} 
\end{figure}

The temperature-dependent magnetic susceptibilities for magnetic fields $\mu_0 H$ applied parallel ($\chi_{\parallel}(T)$) and perpendicular ($\chi_{\perp}(T)$) to the crystallographic $c$-axis for SmAuAl$_4$Ge$_2$ are shown in Fig.~\ref{fig2}. Weak easy $ab$-plane anisotropy is observed in the paramagnetic state, where: (i) $\chi_{\parallel}(T)$ initially decreases with decreasing $T$, evolves through a broad minimum centered near 175 K, and exhibits a gradual increase down to 20 K and (ii) $\chi_{\perp}(T)$ weakly decreases below 300 K, goes through a broad minimum near 250 K, and exhibits a broad maximum centered around 50 K. This behavior is distinct from what is seen for other lanthanides with localized $f$-states, but resembles results for some samarium containing intermetallics and points towards the samarium ions having a trivalent $f$-electron configuration where the $J$ $=$ 5/2 multiplet is strongly impacted by crystal electric field splitting below 300 K.

Antiferromagnetic phase transitions emerge near $T_{\rm{N1}}$ $=$ 13.2 K and $T_{\rm{N2}}$ $=$ 7.4 K. The details of this behavior are shown in Figs.~\ref{fig2}(b) and \ref{fig2}(c), where $\chi_{\parallel}(T)$ is strongly reduced following each transition and weak hysteresis is observed around $T_{\rm{N2}}$. The origin of the hysteresis is not obvious, but we speculate that it indicates the formation of history-dependent magnetic domains within the ordered state. In contrast, $\chi_{\perp}(T)$ weakly increases and then decreases at $T_{\rm{N1}}$ and $T_{\rm{N2}}$, respectively. These trends reveal that the magnetic ordering is characterized by progressively strengthening anti-alignment of the spins along the $c$-axis, with a weak co-alignment of spins at $T_{\rm{N1}}$ and a weak anti-alignment at $T_{\rm{N2}}$ for the in-plane configuration. The occurrence of multiple phase transitions resembles what is seen for $Ln$AuAl$_4$Ge$_2$ analogs that exhibit conventional Curie-Weiss magnetism~\cite{1LnAuAl4Ge2,2LnAuAl4Ge2}, indicates the presence of competing magnetic exchange interactions, and further clarifies how the magnetic anisotropy varies with $Ln$. To further investigate the ordered state, isothermal magnetization measurements were performed at $T$ $=$ 1.8 K. As seen in Fig.~\ref{fig2}(c), $M(H)$ increases linearly with $\mu_0H$ $\leq$ 7 T, with no evidence for metamagnetic phase transitions for both field directions. This contrasts with what is seen for the $Ln$ $=$ Nd, Gd, and Tb analogs, which all show a rich variety of anisotropic metamagnetic phase transitions. Evidence for quantum oscillations is also seen for $H$ $\parallel$ $c$, indicating the high quality of these crystals.

\begin{figure}  \includegraphics[width=\columnwidth]{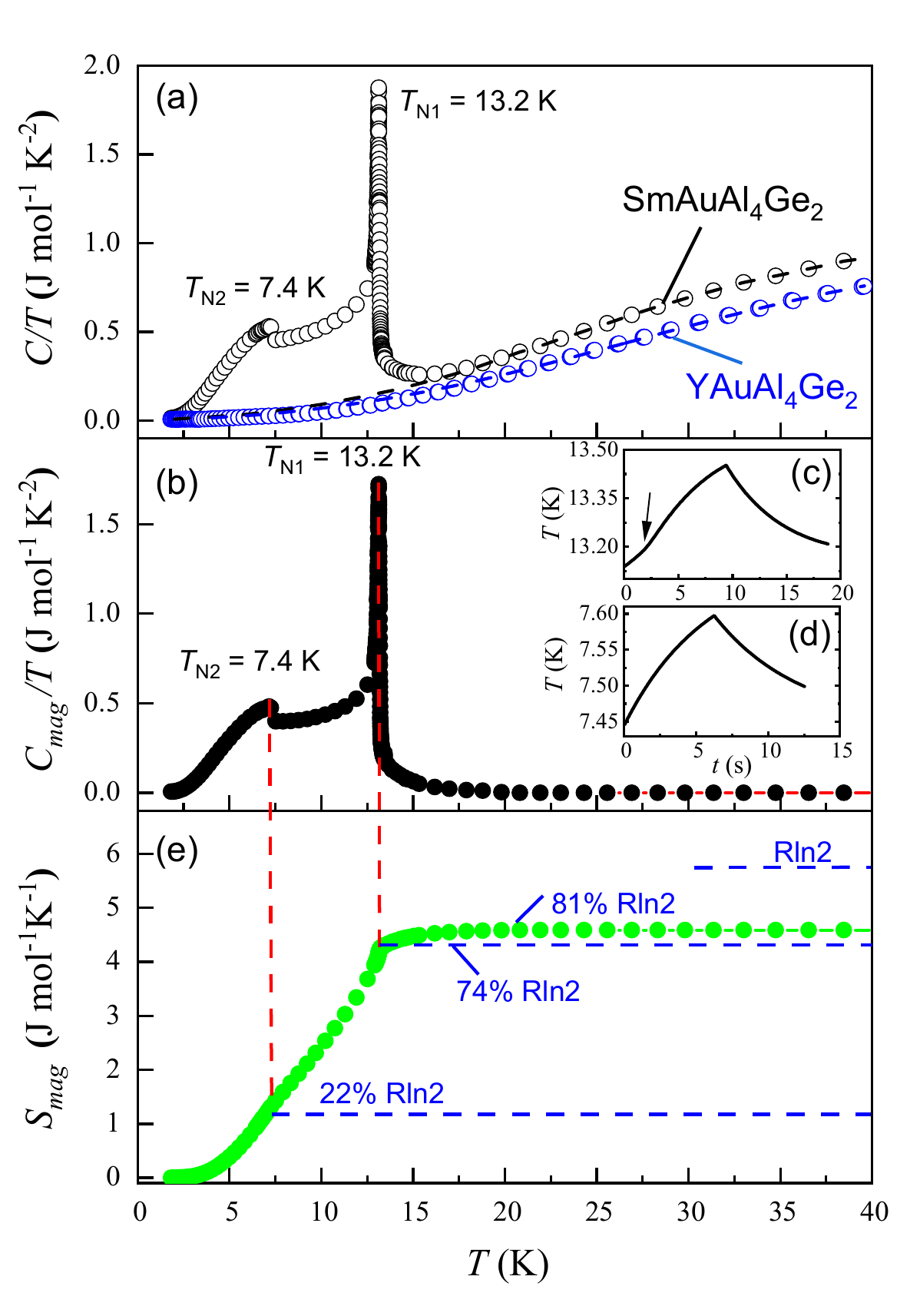}\caption{\label{fig3} (a) The heat capacity divided by temperature $C/T$ vs. $T$ for single-crystal SmAuAl$_4$Ge$_2$ and YAuAl$_4$Ge$_2$ at $T$ = 1.8 - 40 K. The dotted lines represent the fits that are described in the text. (b) The magnetic heat capacity divided by temperature $C_{\rm{mag}}$/$T$ vs. $T$ plotted for SmAuAl$_4$Ge$_2$. (c,d) The relaxation curves around $T_{\rm{N1}}$ (c) and $T_{\rm{N2}}$ (d). In panel (c), the arrow indicates the kink in the heating curve due to the latent heat of the first-order phase transition. A similar feature is not detected around $T_{\rm{N2}}$. (e) Magnetic entropy $S_{\rm{mag}}$ vs. $T$, which is obtained from the heat capacity data as described in the text.}
\end{figure}

The heat capacity divided by temperature $C$/$T$ data for SmAuAl$_4$Ge$_2$ are compared to that of the $J$ = 0 nonmagnetic analog YAuAl$_4$Ge$_2$ in Fig.~\ref{fig3}. As expected, there are qualitative similarities between these curves at elevated temperatures where phonons dominate $C/T$, although the Sm curve lags behind the Y curve with decreasing temperature. Similar behavior is observed in other lanthanide series when the mass of the non-4$f$ ion differs from that of the 4$f$ ion~\cite{GdHC}. To account for this difference, fits to the data (dashed lines) were done using the expression,
\begin{equation}
C(T) = \gamma T + C_{\rm{Debye}}
\end{equation}
where $\gamma$ is the electronic coefficient of the heat capacity and $C_{\rm{Debye}}$ is the Debye integral function. These fits yield $\gamma$ $\approx$ 5 mJ mol$^{-1}$ K$^{-2}$ for both compounds and Debye temperatures $\theta$$_{\rm{D}}$ $=$ 242 K and 220 K for YAuAl$_4$Ge$_2$ and SmAuAl$_4$Ge$_2$, respectively. These $\gamma$ values resemble what was seen for the Gd and Tb analogues~\cite{1LnAuAl4Ge2,2LnAuAl4Ge2}, indicating that the electronic band states are similar, and the primary factors that lead to differences in the heat capacities are the distinct $f$-electron states that are seen for different lanthanides.

There are two pronounced peaks in $C(T)$ at $T_{\rm{N1}}$ and $T_{\rm{N2}}$. The feature at $T_{\rm{N1}}$ is sharp and abrupt, and an examination of the heat pulse relaxation curve $T(t)$ reveals evidence for there being a latent heat, as expected for a first-order phase transition (Fig.~\ref{fig3}(c)). To account for this, in the vicinity of $T_{\rm{N1}}$, the data were analyzed using a single slope expression\cite{PPMSFirstorder}, resulting in the curve shown in Fig.~\ref{fig3}. For the peak near $T_{\rm{N2}}$, the relaxation curves indicate that it is second order (Fig.~\ref{fig3}(d)). To determine the magnetic contribution to the entropy $S_{\rm{mag}}(T)$ (Fig.~\ref{fig3}(e)), we isolate the magnetic contribution to the heat capacity ($C_{\rm{mag}}$/$T$ $=$ $C_{\rm{Sm}}$/$T$ - ($\gamma$$T$ $+$ $C_{\rm{D}}$/$T$)) and then integrate it ($S_{\rm{mag}}(T)$ $=$ $ \int_{0}^{T} C_{\rm{mag}}/T \,dT $). $S_{\rm{mag}}(T$) reaches 4.28 J mol$^{-1}$ K$^{-1}$ at $T_{\rm{N1}}$, which is 74\% of the value expected for a doublet ground state (Rln2) and is strongly reduced from the full $J$ $=$ 5/2 value (Rln6). This is consistent with the perspective from $\chi(T)$ measurements that crystal electric field splitting impacts the low temperature $f$-state behavior. However, we also note that despite the good agreement between $C/T$ and the Debye fit, the finding that $S_{\rm{mag}}$ $<$ Rln2 may imply that phonon background subtraction is an overestimate. As described below, crystal electric field splitting plays an important role over this temperature range, which is not considered in this analysis of the data.

\begin{figure}
   \includegraphics[width=\columnwidth]{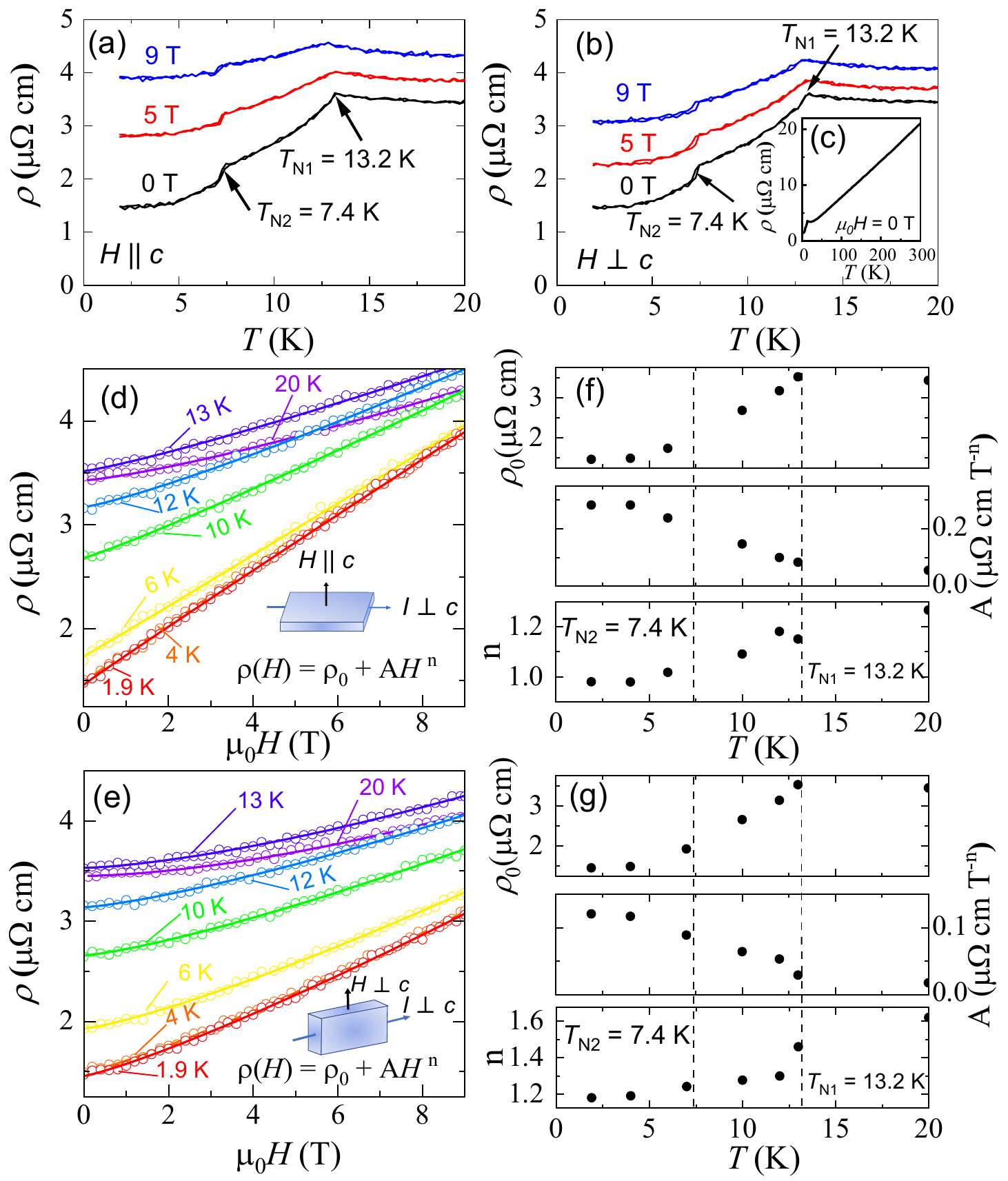}\caption{\label{fig4} (a) Electrical resistivity $\rho(T)$ of SmAuAl$_4$Ge$_2$ near $T_{\rm{N1}}$ and $T_{\rm{N2}}$ shown at different applied fields along the $c$-axis and (b) perpendicular to the $c$-axis.  Data were collected both for zero field cooling and field cooling. (c) $\rho(T)$ for 1.8 $<$ $T$ $<$ 300 K. (d) Isothermal magnetoresistance $\rho(H)$ at different temperatures for magnetic field applied along $c$-axis ($H$ $\parallel$ $c$) and (e) perpendicular to the $c$-axis ($H$ $\perp$ $c$). The electrical current remains in the $ab$ - plane for both cases. The insets show the schematics of the measurement configurations respectively. The data points are represented by open circles and the fits are shown as solid lines.  (f) and (g) The temperature dependence of the parameters $\rho_0$, $A$, and $n$ resulting from fits to the data in (d) and (e), respectively. The dotted vertical lines represent the locations of the ordering temperatures $T_{\rm{N1}}$ and $T_{\rm{N2}}$.}
\end{figure}


Figs.~\ref{fig4}(a)-\ref{fig4}(c) shows the temperature-dependent electrical resistivity $\rho$($T$) with the electrical current $I$ applied in the $ab$ plane and magnetic fields applied either in the $ab$ plane or along the $c$ axis. Consistent with the metallic behavior that is observed for other lanthanide variants~\cite{1LnAuAl4Ge2,2LnAuAl4Ge2}, the room temperature resistivity is near 20 $\mu$$\Omega$ cm and decreases with decreasing $T$. At low temperatures, the magnetic ordering is preceded by a weak minimum that is centered near 20 K. Similar behavior was seen for GdAuAl$_4$Ge$_2$~\cite{1LnAuAl4Ge2}, where it is associated with spin fluctuation scattering. Following this, the transitions at $T_{\rm{N1}}$ and $T_{\rm{N2}}$ both reduce $\rho$($T$) due to the removal of spin disorder scattering. $\rho(T)$ finally saturates towards a value near 1.5 $\mu$$\Omega$ cm at low temperatures, showing that there is little disorder due to crystalline defects. The influence of a magnetic field applied for $H$ $\parallel$ and $\perp$ $c$ are shown in Figs.~\ref{fig4}(a) and ~\ref{fig4}(b). Although the overall value of $\rho(T)$ is enhanced with increasing field, the ordering temperatures remain nearly constant. 

Figs.~\ref{fig4}(d) and \ref{fig4}(e) show the magnetoresistance $\rho(H)$ for electrical current $I$ applied in the $ab$-plane with $H$ $\parallel$ and $\perp$ to the $c$-axis. At the lowest temperatures, $\rho(H)$ is roughly linear over a wide range of fields for $H$ $\parallel$ $c$, with no evidence for metamagnetic phase transitions. $\rho(H)$ subsequently develops weak positive curvature as the temperature is raised through $T_{\rm{N1}}$ and $T_{\rm{N2}}$. As noted previously for GdAuAl$_4$Ge$_2$ and TbAuAl$_4$Ge$_2$ ($I$ $\parallel$ $ab$, $H$ $\parallel$ $c$), this differs from the conventional quadratic magnetoresistance associate with orbital charge carriers' motion and suggests the presence of an unconventional scattering process~\cite{2LnAuAl4Ge2}. In contrast, $\rho(H)$ for $H$ $\perp$ $c$ exhibits positive curvature even at low temperatures, again without evidence for any metamagnetic phase transition. To quantify these behaviors, we carried out fits to the data using a power law expression $\rho(H)$ $=$ $\rho_0$ $+$ $A$$H^n$, which accounts for the residual resistivity $\rho_0$ and the field dependence that arises from the combined electronic and magnetic scattering behaviors. For $H$ $\parallel$ $c$, as shown in Fig.~\ref{fig4} (f), $n$ gradually increases from 1 to 1.3, indicating the persistence of a single dominant scattering mechanism. For $H$ $\perp$ $c$, as shown in Fig.~\ref{fig4} (g), $n$ initially gradually rises from a value near 1.2 until it abruptly increases near $T_{N1}$, presumably as a result of a change changes in the magnetic scattering.


\section{\label{sec:discussion}Discussion}
\begin{figure}
   \includegraphics[width=\columnwidth]{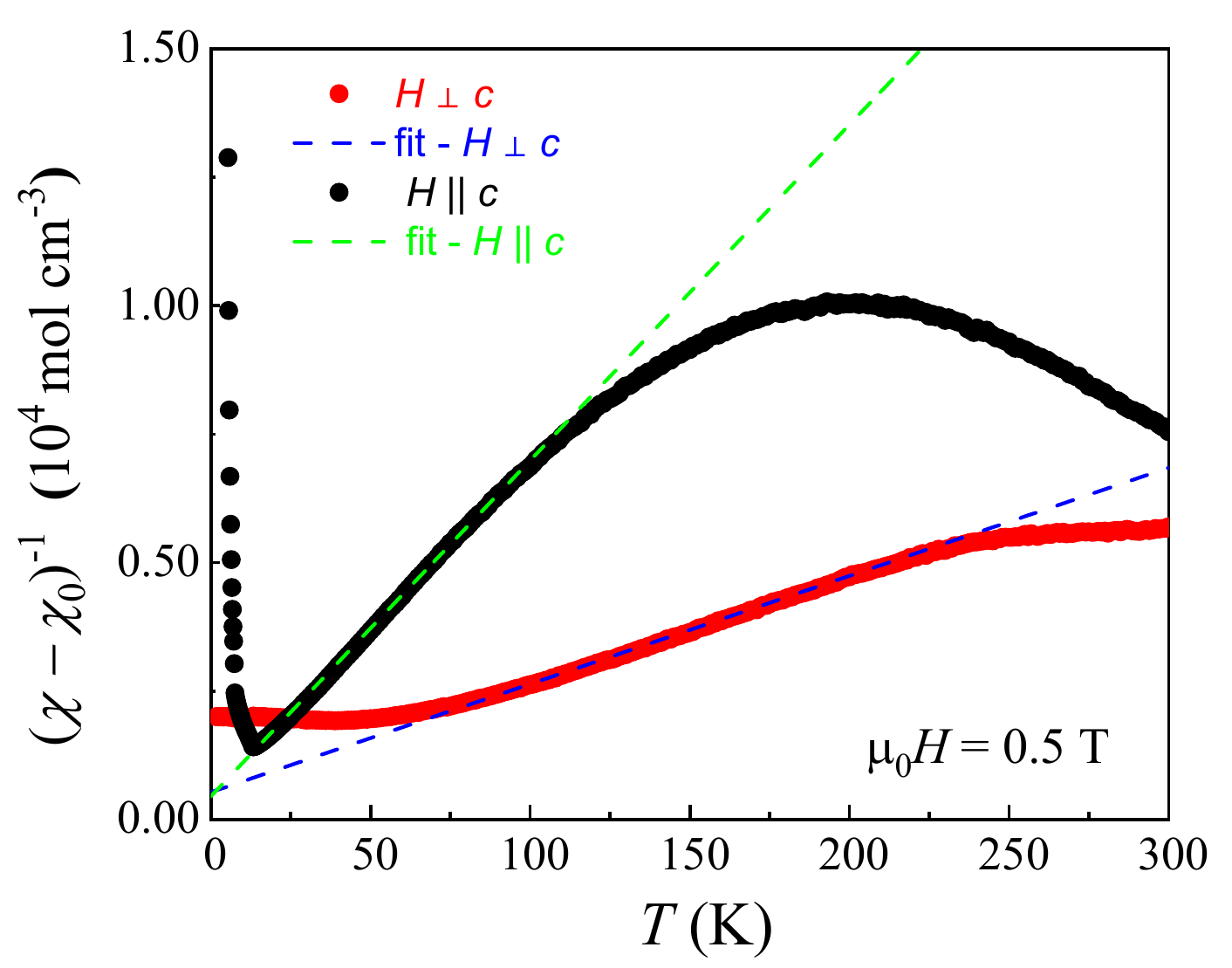}\caption{\label{Fig5} Inverse magnetic susceptibilities ($\chi$ - $\chi_0$)$^{-1}$ for SmAuAl$_4$Ge$_2$ and the fitting results of Eq. (2) (dashed line) are shown for both $H$ $\perp$ $c$ and $H$ $\parallel$ $c$ configuration.}
\end{figure}

These data reveal that SmAuAl$_4$Ge$_2$ exhibits complex magnetic ordering that emerges from a non-Curie-Weiss paramagnetic state. In order to understand this, we first consider a minimal model for the paramagnetism which assumes both that the trivalent $f$-electron state (4$f^5$) dominates the magnetic phenomena and that the crystal electric field (CEF) splitting is sized such that it impacts $\chi(T)$ at and below room temperature. As discussed in Ref.~\cite{CsYbSe2}, the standard Curie Weiss like temperature dependence $\chi(T)$ $=$ $C$/($T-\theta$) is only valid (i) if all of the crystal field split orbitals are homogeneously occupied or (ii) if $T$ $\ll$ $\Delta_0$/$k_{\rm{B}}$ and the quadratic $H$ dependence in the Zeeman splitting is negligible~\cite{CsYbSe2}. Condition ii is unique to 4 $f$ electron systems, where the scale of $\Delta_0$ (the lowest crystal electric field splitting energy) is significantly lower than that of 3 $d$ counterparts. In this situation, the magnetic susceptibility is given by the expression,
\begin{equation}
\chi(T) = \chi_0 + \frac{\nu C_0}{T - \nu \theta} 
\end{equation}
where $C_0$ is the Curie constant ($C_0 = \frac{N_A\mu_0\mu_B^2 g_J^2}{k_B}$), $\theta$ is the Curie-Weiss temperature, and $\nu$ $=$ $g/g_{\rm{J}}$ is a scaled Land\'e $g$-factor $g_J$ = $\frac{2}{7}$). Fits to $\chi (T)$ using Eq.~(2) are shown in Fig.~\ref{Fig5} and the resulting parameters are summarized in Table~\ref{table2}. Importantly, the fits are only expected to work before the first excited state begins to be populated, which is evidenced in the data by the broad maximum that appears in ($\chi_{\parallel} -\chi_0)^{-1}$ near 200 K. Based on this, the lower bound of $\Delta_0$ can be estimated to be on the order of few tens of meV (a few hundreds of Kevin)~\cite{CsYbSe2}. Using the $g$-factor found from the fit, the effective moment $\mu_{\rm{eff}}$ = $g$$\mu_{\rm{B}}$$\sqrt{J(J+1)}$ (where total angular momentum $J$=5/2 was used for Sm$^{3+}$)  is found 0.62 $\mu_{\rm{B}}$ and 0.35 $\mu_{\rm{B}}$ for $H$ ${\parallel}$ and ${\perp}$ to $c$, respectively. These values are slightly reduced from the full Sm$^{3+}$ value of 0.85 $\mu_{\rm{B}}$~\cite{Blundell1}, and the differing values of $\mu_{eff}$ are associated with the anisotropy of the $\nu$ values. Finally, the negative Curie-Weiss temperatures indicate the presence of antiferromagnetic spin exchange interactions.


\begin{table}
\begin{center}
\begin{tabular}{|c|c|c|c|c|c| }\hline
       & $\chi_0$(cm$^3$/mol) & $\theta$(K) & $\mu_{\rm{eff}}$/$\mu_{\rm{B}}$ &$C_0$(emuK/mol) &$\nu$ \\ \hline
 $\chi_{\perp}$ & 9.50 $\times$ 10$^{-4}$ & -16.3 & 0.62 & 3.06 $\times$ 10$^{-2}$ &0.73\\  \hline
 $\chi_{\parallel}$    & 7.05 $\times$ 10$^{-4}$ & -13.8 & 0.35 &3.06 $\times$ 10$^{-2}$ &0.41\\ \hline
\end{tabular}
\caption{\label{table2} Summary of values obtained from fits to the magnetic susceptibility $\chi(T)$ data using Eqn. 2.}
\end{center}
\end{table}

While this picture qualitatively describes the data, there also are several features that deviate from it. First, we note that 1/$\chi_{\perp}(T)$ is strongly suppressed from the fit values below 70 K. The reason for this is not clear, but we speculate that it might be attributed to anisotropic spin fluctuations in the vicinity of ordering temperature: e.g., relating to spin frustration in the $ab$ plane. It is also important to note that different minima in $\chi(T)$ are observed for the two field directions. This is unexpected since CEF splitting in zero field is isotropic, and thus should impact both curves in similar ways. This implies that additional effects (e.g., van Vleck splitting between the $J$ $=$ 5/2 and 7/2 states) may need to be considered to fully understand this unusual paramagnetic state. A spectroscopy study such as inelastic neutron scattering will be useful to render an accurate CEF characterization.


\begin{table*}
\centering
\begin{tabular}{||c||c c c c c||}
\hline \hline
~~~~&~~~~Ce\cite{CeAuAl4Ge2}~~~~~~
                  &~~~~Nd\cite{3NdAuAl4Ge2.PRM.2023}~~
                  &~~~~Sm~~~~ &~~~~Gd\cite{1LnAuAl4Ge2}~~~~
                  &~~~~Tb\cite{1LnAuAl4Ge2}~~~~\\           
\hline \hline
$T_{\rm{N1}}$(K) - ZFC ~&1.40 
                  &1.75
                  &13.2
                  &17.8
                  &13.9\\ 
\hline                  
$T_{\rm{N2}}$(K) - ZFC~&- 
                  &0.49
                  &7.4
                  &15.6
                  &9.8\\                     
\hline
$T_{\rm{N3}}$(K) - ZFC~&- 
                  &-
                  &-
                  &13.8
                  &- \\                   
\hline
Anisotropy        &0.3 
                  & 1.62
                  & 2.53
                  & 1.56
                  & 5.9\\ 
$\chi_{\parallel}$/$\chi_{\perp}$ at 1.8K ~& 
                  &
                  &
                  &
                  &\\
\hline
$H_{\rm{C1}}$ (T)~& -
                  & 0.04($H \parallel c$)
                  & -
                  & 1.9($H \perp c$)
                  & 1.3($H \perp c$)\\
\hline
$H_{\rm{C2}}$ (T)~& -
                  & 0.75($H \parallel c$)
                  & -
                  & -
                  & 1.9($H \perp c$)\\
\hline
$H_{\rm{C3}}$ (T)~& -
                  & 1.6 ($H \parallel c$)
                  & -
                  & -
                  & 2.7($H \perp c$)\\
\hline
$M_{\perp}$ ($\rm{\mu_B}$) &1.3 
                  &1.23
                  & 1.74 $\times 10^{-2}$
                  & 4.9
                  &8.31\\ 
\hline                  
$M_{\parallel}$ ($\rm{\mu_B}$) &0.4 
                  &1.41
                  & 6.7 $\times 10^{-3}$
                  &3.65
                  &1.33\\ 
                             
\hline
$S_{mag}$($T$ = $T_{\rm{N1}}$) &3.00 
                  &3.46
                  &4.28 
                  &12.8
                  &13.9\\
(J/(mol-K))~& 
                  &
                  &
                  &
                  &\\
\hline                  
$S_{mag}$($T$ = 70K)   & - 
                  &14
                  &0.56
                  &16.4
                  &20.7\\ 
(J/(mol-K))~&
                  &
                  &
                  &
                  &\\                   
\hline  
$R$ln(2$J$ + 1)   & 14.9
                  &19.1
                  &14.9
                  &17.3
                  &21.3\\ 
(J/(mol-K))        ~&
                  &
                  &
                  &
                  &\\                    
\hline  \hline                  
\end{tabular}
\caption{\label{table1} Summary of magnetic properties for $Ln$AuAl$_4$Ge$_2$ ($Ln$ $=$ lanthanide) obtained from the magnetic susceptibility $\chi$($T$), magnetization $M(H)$, and heat capacity $C(T)$, where $\chi(T)$ was collected at $\mu_0H$ $=$ 0.5 T and $M$ values here are reported at 7 T and 1.8 K. Magnetic entropy $S_{\rm{mag}}(T)$ are calculated from $C_{\rm{mag}}(T)$ (See text). $T_{\rm{N1}}$, $T_{\rm{N2}}$, and $T_{\rm{N3}}$ refers the zero field ordering temperatures identified from $\chi$($T$). $\perp$ ($\parallel$) refers to $H \perp c$ ($H \parallel c$). $R$ refers the gas constant and $J$ is the total angular momentum for each Ln$^{3+}$ ions. Data for $Ln$ = Ce, Nd, Gd, and Tb are from Refs~\cite{CeAuAl4Ge2,1LnAuAl4Ge2,2LnAuAl4Ge2,3NdAuAl4Ge2.PRM.2023}.}
\end{table*}

\begin{figure}
   \includegraphics[width=\columnwidth]{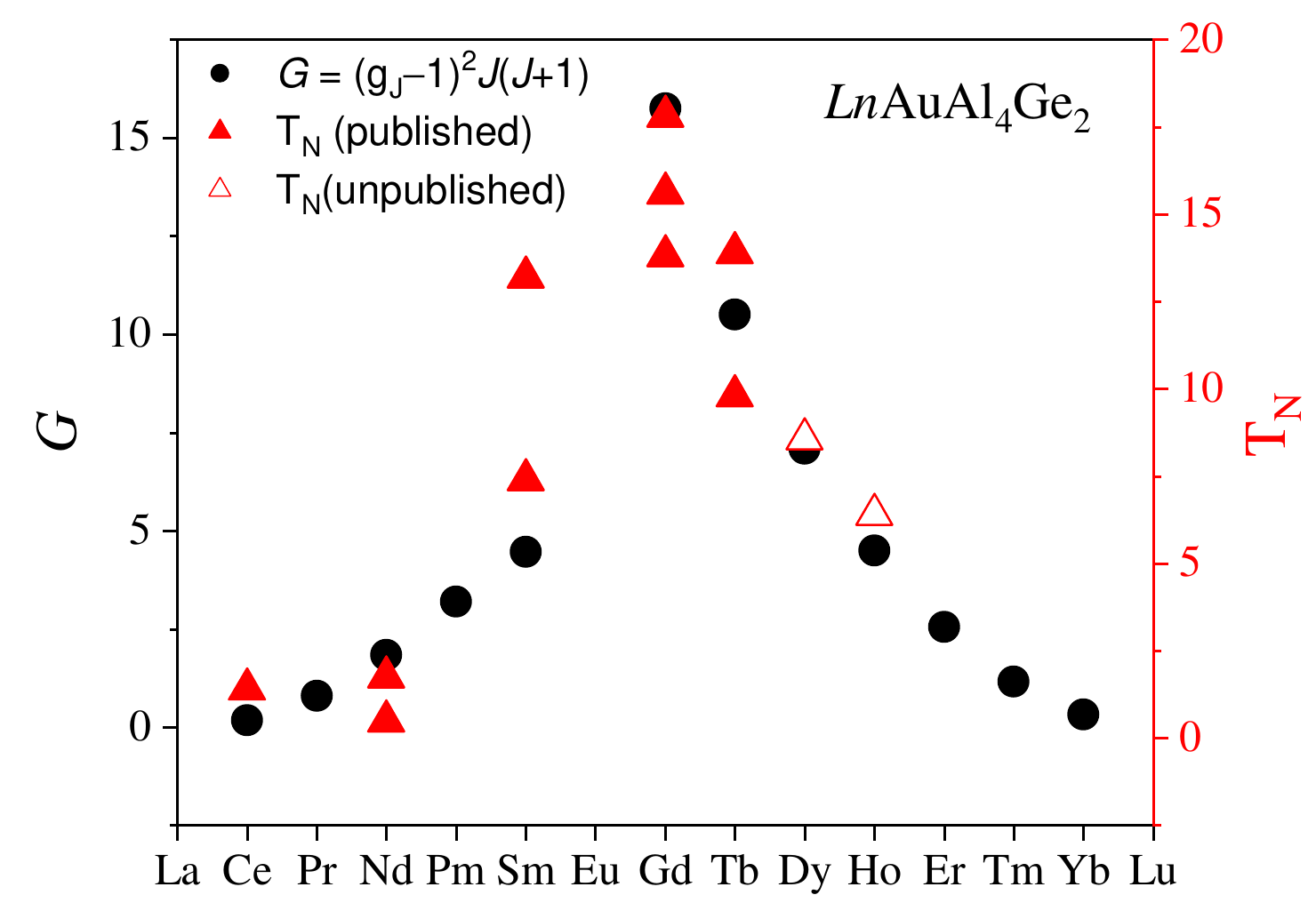}\caption{\label{fig6} Comparison between the de Gennes scaling factor $G$ (defined in the text) and the ordering temperatures $T_{\rm{N1}}$ and $T_{\rm{N2}}$ (triangles) for the $Ln$AuAl$_4$Ge$_2$ series. Data for $Ln$ $=$ Ce, Nd, Gd, and Tb are from Refs.~\cite{CeAuAl4Ge2,1LnAuAl4Ge2,2LnAuAl4Ge2,3NdAuAl4Ge2.PRM.2023}. Empty triangles are from unpublished data~\cite{unpublished}.}
\end{figure}

This behavior contrasts with what is seen for other $Ln$AuAl$_4$Ge$_2$ analogues with conventional Curie-Weiss paramagnetism (Table~\ref{table1}), and might lead to the expectation that the ground state would show distinct behavior. Despite this, there are noteworthy similarities in the ordered states between SmAuAl$_4$Ge$_2$ and its analogs. This is highlighted in Fig.~\ref{fig6}, where we plot the ordering temperatures for several examples ($Ln$ $=$ Ce, Nd, Sm, Gd, Tb, and Dy) and the de Gennes scaling factor ($G$ $=$ (g$_{\rm{J}}$-1)$^2$$J$($J$ $+$ 1)) vs. lanthanide~\cite{deGennes1,deGennes2}. From this, it is clear that there is close agreement between $G$ and the trends seen in the ordering temperatures, indicating that these compounds have a shared spin exchange mechanism. An explanation for this could be that while each of these compounds has a different CEF splitting for the $f$-electron state, which impacts the effective magnetic moment, they all share similar Fermi surface topographies. This would result in them having similar conduction electron-mediated RKKY spin interactions that are robust against variations of the lanthanide ion. Indeed, this is expected in the absence of hybridization between the $f$- and conduction electron.

Another important distinction between SmAuAl$_4$Ge$_2$ and its relatives is the lack of metamagnetic phase transitions. This is puzzling, but given that the $M(H)$ curves for SmAuAl$_4$Ge$_2$ are non-saturating at 7 T, one possibility is that they will appear at larger fields than those accessed in this study. The reason for this might be that the relatively small Sm magnetic moment modifies the internal field in a way that enhances the field-driven transition energy scale. Alternatively, it is possible that both measured directions represent hard magnetic axes. More detailed measurements (e.g., exploring the in-plane magnetic anisotropy) will be useful to address this question. 

It is also appealing to consider that the RKKY interaction alone may not fully account for the complex magnetic order. All of these compounds, with the exception of CeAuAl$_4$Ge$_2$, exhibit multiple temperature and field-dependent transitions, indicating the presence of magnetic frustration. This motivates the need for further work to understand effects arising from (i) competing RKKY interactions: e.g., as seen for the anisotropic next nearest neighbor Ising model\cite{ANNNI.model.1988} and (ii) geometric frustration. In any case, these behaviors open an intriguing path for stabilizing and tuning nontrivial spin states: e.g., similar to what is seen for the structurally similar Gd$_2$PdSi$_3$~\cite{Gd2PdSi3}, where a unique combination of spin anisotropy with a centrosymmetric triangular Gd lattice produces an unusual skyrmion state. For the $Ln$AuAl$_4$Ge$_2$ compounds, if skyrmions or other nontrivial magnetic textures do not appear in the $T-H$ phase diagram of one of the parent compounds, it may be possible to access them by chemically mixing the $f$-element site. This would preserve the structural constraints but vary the magnetic anisotropy, magnetic moment, and the resulting $T-H$ phase diagrams.


Finally, it is intriguing that linear magnetoresistance is seen for fields applied along $c$-axis. This is similar to what is seen for the Gd and Tb analogues~\cite{2LnAuAl4Ge2}, and based on this commonality we infer that the underlying origin for this behavior is independent of the details of the $f$-electron state. This naturally leads us to consider that it connects to electronic degrees of freedom that are shared amongst all of the chemical analogues. As we noted previously, a possible scenario is that the formation of charge or spin density waves leads to modification of high curvature Fermi surfaces due to zone-folding energy gaps - thereby producing linear magnetoresistance~\cite{Kolincio2020,Tsuda2018,FengY2019}. More recently, angle-respolved photoemission spectroscopy measurements also revealed the presence of nontrivial topologically protected bands~\cite{GdAuAl4Ge2-2023}, whose impact on the electrical transport remains to be clarified. In order to investigate these points, it will be of interest to measure the magnetoresistance of the Y, Ce, and Nd analogs, to search for Fermi surface similarities or instabilities (e.g., by detecting quantum oscillations), and to perform measurements to even higher fields to reveal the extent of linear magnetoresistance throughout the entire family. 

\section{\label{sec:conclusions}Conclusions}
In summary, we have shown that SmAuAl$_4$Ge$_2$ exhibits weak paramagnetism that strongly deviates from conventional Curie-Weiss behavior. This is described in terms of crystal electric field splitting, where the energy difference between the ground state and the first excited state is on the order of several tens of meV. Similar to other $Ln$AuAl$_4$Ge$_2$ analogues, complex antiferromagnetic ordering emerges at low temperatures ($T_{\rm{N1}}$ $=$ 13.2 K and $T_{\rm{N2}}$ $=$ 7.4 K). This behavior likely relates to the geometrically frustrated triangular arrangement of $Ln$ ions in the $ab$ plane, but other factors such as complexity in the RKKY interaction may play an important role. The temperature-dependent electrical resistivity and heat capacity indicate standard metallic behavior, although linear magnetoresistance appears within the ordered states over a wide field range. Taken together, this low temperature behavior resembles what is seen for other $Ln$AuAl$_4$Ge$_2$ materials, with some noteworthy differences including the lack of metamagnetic phase transitions. Thus, SmAuAl$_4$Ge$_2$ emerges as an environment for complex quantum spin states and unusual magnetotransport behaviors and invites further investigations of the entire family of materials. In particular, it will be useful to measure the order parameters (e.g., using neutron scattering), to quantify the in-plane magnetic anisotropy, to search for metamagnetic phase transitions at even larger magnetic fields, and to develop a better understanding of the Fermi surface topography and possible topology (e.g., using quantum oscillations or angle-resolved photoemission spectroscopy measurements.)

\section{\label{sec:Acknowledgements}Acknowledgements}
RB, KF, and OO were supported by the National Science Foundation through NSF DMR-1904361. CB was supported by the National High Magnetic Field Laboratory Research Experience for Undergraduates program. M.L. was supported by the U.S. Department of Energy, Basic Energy Sciences, Materials Sciences and Engineering Division under Award No. DE-SC0021377.The National High Magnetic Field Laboratory is supported by the National Science Foundation through NSF DMR-1644779 and the State of Florida.

\end{document}